\documentclass[12pt,a4paper]{article}
\usepackage{a4wide}
\usepackage{latexsym}
\usepackage{epsf}
\usepackage{amssymb}
\begin{document}
\def\be{\begin{equation}}
\def\ee{\end{equation}}
\def\bea{\begin{eqnarray}}
\def\eea{\end{eqnarray}}

\def\pd{\partial}
\def\a{\alpha}
\def\b{\beta}
\def\g{\gamma}
\def\d{\delta}
\def\m{\mu}
\def\n{\nu}
\def\t{\tau}
\def\l{\lambda}
\def\O{\Omega}
\def\r{\rho}
\def\s{\sigma}
\def\e{\epsilon}
\def\scri{\mathcal{J}}
\def\cM{\mathcal{M}}
\def\tcM{\tilde{\mathcal{M}}}
\def\RR{\mathbb{R}}

\hyphenation{re-pa-ra-me-tri-za-tion}
\hyphenation{trans-for-ma-tions}


\begin{flushright}
IFT-UAM/CSIC-03-23\\
gr-qc/0307090\\
\end{flushright}

\vspace{1cm}

\begin{center}

{\bf{\Large Loops versus strings}\footnote{Talk given in the meeting {\em What comes beyond 
the standard model?}, Portoroz, Eslovenia, 12-17 July 2003.}}

\vspace{.5cm}

{\bf Enrique \'Alvarez }

\vspace{.3cm}

\vskip 0.4cm

{\it  Instituto de F\'{\i}sica Te\'orica UAM/CSIC, C-XVI,
and  Departamento de F\'{\i}sica Te\'orica, C-XI,\\
  Universidad Aut\'onoma de Madrid 
  E-28049-Madrid, Spain }

\vskip 0.2cm

\vskip 1cm

{\bf Abstract}

\end{center}

\begin{quote}
Two popular attempts to understand the quantum physics of gravitation
are critically assessed. The talk on which this paper is based 
was intended for a general particle-physics audience.  
  
\end{quote}


\newpage

\setcounter{page}{1}
\setcounter{footnote}{1}
\newpage
\section{General Questions on Quantum Gravity}
It is not clear at all what is the problem in quantum gravity (cf. \cite{Alvarez} or 
\cite{Alvarez-Gaume} for 
general reviews, written in the same spirit as the present one).
The answers to the following questions are not known, and I believe it can do no harm
to think about them before embarking in a more technical discussion.
\par
Actually, some people  proposed that gravity should not be quantized, owing to its
special properties as determining the background on which all other fields propagate.
There is a whole line of thought on the possibility that gravity is not a fundamental theory,
and this is certainly an alternative  one has to bear in mind. Indeed, even the {\em holographic
principle} of G. 't Hooft, to be discussed later, can be interpreted in this sense.
\par
Granting that, the next question is whether 
it does  make any sense to consider gravitons
  propagating in some background; that is, whether there is some useful
approximation in which there is a particle physics approach to the physics
of gravitons as quanta of the gravitational field. A related question is whether 
semiclassical gravity, i.e., the approximation
in which the source of the classical Einstein equations is replaced by the
expectation value of the energy momentum tensor of some quantum theory has some physical 
(\cite{Duff}) validity
in some limit.
\par
At any rate, even if it is possible at all, the at first sight easy problem of graviton interactions
in an otherwise flat background has withstood analysis of several generations of physicists.
The reason is that the coupling constant has mass dimension $-1$, so that the structure
of the perturbative counterterms involve higher and higher orders in the curvature invariants
(powers of the Riemann tensor in all possible independent contractions), schematically,
\be
S=\frac{1}{2\kappa_R^2}\int R+\int R^2 + \kappa_R^2 \int R^4+\ldots
\ee
Nobody knows how to make sense of this approach, except in one case, to be mentioned later on.
\par
It could be possible, {\em sensu stricto} to stop here. But if we believe that 
quantum gravity should  give answers to such questions as to the fate of the initial 
cosmological singularity, its is almost unavoidable to speak of the wave function of the universe.
This brings its own set of problems, such as to whether it is possible to do quantum mechanics
 without classical observers or whether  the wave function of the Universe
 has a probablilistic interpretation. Paraphrasing C. Isham \cite{Isham}, one would not known when to 
qualify a probabilistic prediction on the whole Universe as a successful one.
\par
The aim of the present paper is to discuss in some detail established results on the field.
In some strong sense, the review could be finished at once, because there are none. 
There are, nevertheless, some interesting attempts, which look promising from certain points of view.
Perhaps the two approaches that have attracted more attention have been the loop approach,
on the one hand and strings on the other. We shall try to critically assess prospects in both.
 Interesting related papers are \cite{Horowitz}\cite{Smolin}.

\section{The issue of background independence}
One of the main differences between both attacks to the quantum gravity problem
 is the issue of background
 independence, by which it is understood that no particular background should enter
into the definition of the theory itself. Any other approach is purportedly at
variance with diffeomorphism invariance.
\par
Work in particle physics in the second half of last century led to some
understanding of ordinary gauge theories. Can we draw some lessons from there?
\par
Gauge theories can be formulated in the {\em bakground field approach}, as
introduced by B. de Witt and others (cf. \cite{dewitt}). In this approach,
the quantum field theory depends on a background field, but not on any one in particular,
and the theory enjoys background gauge invariance.
\par
Is it enough to have quantum gravity formulated in such a way? This was, incidentally, the way
G.  Hooft and M. Veltman did the first complete one-loop calculation (\cite{thv}).
\par
It can be argued that the only 
vacuum expectation value consistent with diffeomorphisms invariance is
\be
<0|g_{\a\b}|0>=0
\ee
in which case the answer to the above question ought to be in the negative, because this is a singular
background and curvature invariants do not make sense.
It all boils down as to whether the ground state of the theory is diffeomorphisms invariant or not.
There is an example, namely three-dimensional gravity in which invariant quantization can be performed
\cite{Witten3}. In this case at least, the ensuing theory is almost topological. 
\par
In all attempts of a canonical quantization of the gravitational field, 
one always ends up with an (constraint) equation
corresponding physically to the fact that the total hamiltonian of a parametrization invariant theory
should vanish. When expressed in the Schr\"odinger picture, this equation is often 
dubbed the {\em Wheeler-de Witt equation}. This equation is plagued by
operator ordering and all other sorts of ambiguities.
It is curious to notice that in ordinary quantum field theory there also exists a Schr\"odinger
representation, which came recently to be controlled well enough as to be able to perform lattice 
computations (\cite{Luscher}).
\par
Gauge theories can be expressed in terms of gauge invariant operators, such as
Wilson loops . They obey a complicated
set of equations, the loop equations, which close in the large $N$ limit
as has been shown by Makeenko and Migdal (\cite{MM}). These equations can be properly regularized,
e.g. in the lattice. Their explicit solution
is one of the outstanding challenges in theoretical physics. Although many conjectures
have been advanced in this direction, no definitive result is available.
\section{ Loops}
The whole philosophy of this approach is canonical, i.e., an analysis of the evolution of
variables defined classically through a foliation of spacetime by a family of spacelike three-surfaces
$\Sigma_t$. The standard choice in this case (cf. for example \cite{Alvarez}) is the 
three-dimensional metric, $g_{ij}$, and its canonical conjugate, related to the extrinsic curvature.
Due to the fact that the system is reparametrization invariant, the total hamiltonian vanishes,
and this hamiltonian constraint is usually called the Wheeler- de Witt equation.
\par
Here, as in any canonical approach the way one chooses the canonical variables is fundamental.
\par
Ashtekar's clever insight started from the definition of an original set of 
variables (\cite{Ashtekar}) stemming from the Einstein-Hilbert 
lagrangian written in the form \footnote{Boundary terms have to be considered as well. We refer to 
the references for
details.}
 \be
S=\int e^a\wedge e^b\wedge R^{cd}\epsilon_{abcd}  
\ee
where $e^a$ are the one-forms associated to the tetrad,
\be
e^a\equiv e^a_{\m}dx^{\m}.
\ee
Tetrads are defined up to a local Lorentz transformation
\be
(e^a)^{\prime}\equiv L^a\,_b(x)e^b
\ee
The associated $SO(1,3)$ connection one-form $\omega^a\,_b$ is usually called the 
{\em spin connection}. Its field strength is the  curvature expressed as a two form:
\be
R^a\,_b\equiv d\omega^a\,_b+\omega^a\,_c\wedge \omega^c\,_b.
\ee
Ashtekar's variables are actually based on the $SU(2)$ self-dual connection
 \be
 A=\omega - i * \omega
\ee
Its field strength is
\be
F\equiv d A + A\wedge A
\ee
The dynamical variables are then $(A_i, E^i\equiv F^{0i})$. The main virtue of these variables is that
constraints are then linearized.
One of them is exactly analogous to            
Gauss'law: 
\be
D_i E^i=0.
\ee
There is another one related to three-dimensional diffeomorphisms invariance,
\be
tr\, F_{ij}E^i=0
\ee
 and, finally, there is the Hamiltonian constraint,
\be 
 tr F_{ij}E^i E^j=0
\ee
\par
On a purely mathematical basis, there is no doubt that Astekhar's variables are of a great ingenuity.
As a physical tool to describe the metric of space, they are not real in general. This forces
a reality condition to be imposed, which is akward. For this reason it is usually prefered
to use the Barbero-Immirzi (\cite{Barbero}\cite{Immirzi}) 
formalism in which the connexion depends on a free parameter, $\gamma$,
\be
A_a^i=\omega_a^i +\gamma K_a^i
\ee
$\omega$ being the spin connection and $K$ the extrinsic curvature. When $\gamma=i$ Astekhar's
formalism is recovered; for other values of $\gamma$ the explicit form of the constraints 
is more complicated. Thiemann (\cite{Thiemann}) has proposed a form for the Hamiltonian constraint
which seems promising, although it is not clear whether the quantum constraint 
algebra is isomorphic to the classical algebra (cf.\cite{Rovellir}). A comprehensive reference
is \cite{Thiemannr}
\par
Some states which satisfy the Astekhar constraints are given by the
 loop representation, which can be introduced from the construct (depending both on
the gauge field $A$ and on a parametrized loop $\gamma$)
\be
 W (\gamma , A)\equiv tr\, P e^{\oint_{\gamma}A}
\ee
and a functional transform mapping functionals of the gauge field $\psi(A)$ into functionals
of loops, $\psi(\gamma)$:
\be
 \psi(\gamma)\equiv \int {\cal D}A\, W(\gamma,A)\psi(A)
\ee
When one divides by diffeomorphisms, it is found that
functions of knot classes (diffeomorphisms classes of smooth, non self-intersecting loops)
 satisfy all the constraints. 
 \par
Some particular states sought to reproduce smooth spaces at coarse graining are
the {\em Weaves}. It is not clear to what extent they also approach the conjugate variables (
that is, the extrinsic curvature) as well.
\par
In the presence of a cosmological constant the hamiltonian constraint reads:
\be
\epsilon_{ijk}E^{ai}E^{bj}(F^k_{ab}+\frac{\lambda}{3}\epsilon_{abc}E^{ck})=0
\ee
A particular class of solutions of the constraint \cite{Smolinc} are self-dual solutions of the form
\be
F^i_{ab}=-\frac{\lambda}{3}\epsilon_{abc}E^{ci}
\ee
Kodama (\cite{Kodama} has shown that the Chern-Simons state 
 \be
 \psi_{CS}(A)\equiv e^{\frac{3}{2\lambda} S_{CS}(A)}
\ee 
is a solution of the hamiltonian constraint. He even suggested that the {\em sign} of the coarse
grained, classical  cosmological constant was always positive, irrespectively of the sign of the 
quantum 
parameter $\lambda$, but it is not clear whether this result is general enough.
 There is some concern that this state as such is not normalizable with the usual norm. It 
has been argued that
this is only natural, because the physical relevant norm must be very different from the na\"{\i}ve one
(cf. \cite{Smolin}) and indeed normalizability of the Kodama state has been suggested as a criterion
for the correctness of the physical scalar product.
\par
Loop states in general (suitable symmetrized) can be represented as 
spin network (\cite{RS}) states: colored lines (carrying some $SU(2)$ representation) 
meeting at nodes where intertwining $SU(2)$ operators act. A beautiful graphical
representation of the group theory has been succesfully adapted for this purpose.
There is a clear relationship between this representation and the Turaev-Viro \cite{Turaev} 
invariants
Many of these ideas have been foresighted by Penrose (cf. \cite{penrose}).
\par
There is also a 
path integral representation, known as  {\em spin foam} (cf.\cite{Baez}), a topological theory 
of colored surfaces representing the evolution of a spin network.
These are closely related to topological BF theories, and indeed, independent generalizations 
have been proposed.
Spin foams can also be considered as an independent approach to the quantization of the gravitational 
field.(\cite{Barrett})
\par
In addition to its specific problems, this approach shares with all canonical approaches
to covariant systems
the problem of time. It is not clear its definition, at least in the absence of matter.
Dynamics remains somewhat mysterious; the hamiltonian constraint does not say in what sense
(with respect to what) the three-dimensional dynamics evolve.

\subsection{Big results}
One of the main successes of the loop approach is that the
area (as well as the volume) operator is discrete. This allows, assuming that 
a black hole has been formed (which is a process that no one knows how to represent
in this setting), to explain the formula for the black hole entropy . The result
is expressed in terms of the Barbero-Immirzi parameter (\cite{RSS}). The physical meaning
 of this dependence is not well understood.
\par
 
It has been pointed out \cite{Bekenstein} that there  is a potential drawback
in all theories in which the area (or mass) spectrum is discrete with eigenvalues $A_n$ 
if the level spacing between eigenvalues $\d A_n$ is uniform because of the predicted thermal character
of Hawking's radiation. The explicit computations in the present setting, however, lead to
an space between (dimensionless) eigenvalues
\be
 \delta\, A_n\sim e^{-\sqrt{A_n}}, 
 \ee 
which seemingly avoids this set of problems.
  \par 
It has also been pointed out that \cite{Freidel} not only the spin foam, but almost
all other theories of gravity can be expressed as topological BF theories with constraints.
While this is undoubtely an intesting and potentially useful remark, it is
important to remember that the difference between the linear sigma model (a free field theory)
and the nonlinear sigma models is just a matter of constraints. This is enough to produce a mass gap
and asymptotic freedom in appropiate circumstances.

\section{Strings}
It should be clear by now that we probably still do not know
what is exactly the problem to which string theories are the answer. At any rate, the starting 
point is that all elementary particles are viewed as quantized excitations of a one dimensional object,
the string, which can be either open (free ends) or closed (a loop). Excellent books are 
avaliable, such as \cite{Greens}\cite{Polchinskis}.
\par
String theories enjoy a convoluted history. Their origin can be traced to the
Veneziano model  of strong interactions. A crucial step was the reinterpretation 
by Scherk and Schwarz 
(\cite{Scherk}) of the massless spin two state in the closed sector (previously 
thought to be related to the 
Pomeron) as the  
graviton and consequently of the whole string theory as
a potential 
theory of quantum gravity, and potential unified theories of all interactions. Now the wheel has
made a complete turn, and we are perhaps back
through the Maldacena conjecture (\cite{Maldacena}) to a closer relationship than previously thought 
with ordinary gauge theories.
\par
From a certain point of view, their dymamics is determined by a two-dimensional non-linear sigma
model, which geometrically is a theory of imbeddings of a two-dimensional surface (the world
sheet of the string) to a (usually ten-dimensional) target space:
\be
x^{\m}(\xi): \Sigma_2\rightarrow M_n
\ee
There are two types of interactions to consider.
Sigma model interactions (in a given two-dimensional surface) 
are defined as an expansion in powers of momentum, where a new
dimensionful parameter, $\a^{\prime}\equiv l_s^2$ sets the scale. This scale
is {\em a priori} believed to be of the order of the Planck length. The first terms in the action
always include a coupling to the massless backgrounds: the spacetime metric, the two-index Maxwell
like field known as the Kalb-Ramond or $b$-field, and the dilaton. To be specific,
\be
 S= \frac{1}{l_s^2}\int_{\Sigma_2}g_{\mu\nu}(x(\xi))\partial_a x^{\mu}(\xi)\partial_b 
x^{\nu}(\xi)\gamma^{ab}(\xi)
+\ldots 
\ee 
There are also string interactions, (changing the two-dimensional surface) 
proportional to the string coupling constant, $g_s$, whose
variations are related to the logarithmic variations of the dilaton field.
Open strings (which have gluons in their spectrum) {\em always} contain closed strings 
(which have gravitons in their spectrum) as intermediate
states in higher string order ($g_s$)  corrections. This interplay open/closed is one of the 
most fascinating aspects of the whole string theory.
\par
It has been discovered by Friedan (cf. \cite{Friedan}) that in order for the quantum theory
to be consistent with all classical symmetries (diffeomorphisms and conformal invariance), the 
beta function of
the generalized couplings \footnote{There are corrections coming from both dilaton and Kalb-Ramond
fields. The quoted result is the first term in an expansion in derivatives, with expansion parameter
 $\a^{\prime}\equiv l_s^2$.} must vanish:
\be
 \beta (g_{\mu\nu})=R_{\mu\nu}=0 
\ee 
  This result remains until now as one of the most important ones in string theory, hinting at a 
deep relationship between Einstein's equations and the renormalization group.
   
Polyakov (\cite{Polyakov}) introduced the so called {\em non-critical strings} which have
in general a two-dimensional cosmological constant (forbidden otherwise by Weyl invariance).
The dynamics of the conformal mode (often called Liouville in this context) is, however, 
poorly understood.

\subsection{General setup}

Fundamental strings live in  D=10 spacetime dimensions, and so a Kaluza-Klein mecanism of sorts
must be at work in order to explain why we only see four non-compact dimensions at low energies.
Strings have in general tachyons in their spectrum, and the only way to construct seemingly
consistent string theories (cf. \cite{Gliozzi}) 
is to project out those states, which leads to supersymmetry. This means in turn that all low energy
predictions heavily depend on the supersymmetry breaking mechanisms. 
\par
  
 String perturbation theory is probably well defined although a full proof is not available.
\par
Several stringy symmetries are believed to be exact: T-duality, relating large and small compactification
volumes, and $S$-duality,
relating the strong coupling regime with  the weak coupling one.
Besides, extended configurations ({\em D branes}); topological defects in which open strings can
end are known to be important \cite{Polchinski}. They couple to Maxwell-like fields which are p-forms
called Ramond-Ramond (RR) fields.
These dualities \cite{Hull} relate  all five string theories (namely, Heterotic $E(8)\times E(8)$ 
Heterotic $SO(32)$, Tipe $I$, $IIA$ and $IIB$) and it is conjectured that there is an 
unified eleven -dimensional theory, dubbed $M$-theory
of which ${\cal N}=1$ supergravity in $d=11$ dimensions is the low energy limit.

\subsection{Big results}
Perhaps the main result is that graviton physics in flat space is well defined for the first time, and
this is no minor accomplishment.
\par
Besides, there is evidence that at least some geometric singularities are harmless
in the sense that strings do not feel them.
Topology change amplitudes do not vanish in string theory.
\par
The other Big Result \cite{Strominger} is that one can
correctly count states
 of extremal black holes
 as a function of charges. This is at the same time astonishing and disappointing.
It clearly depends strongly on the objets being BPS states (that is, on supersymmetry),
and the result has not been extended to non-supersymmetric configurations.
On the other hand, as we have said, it {\em exactly} reproduces the entropy as a 
function of a sometimes large number of charges, without any adjustable parameter.
     
\subsection{The Maldacena conjecture}
 Maldacena \cite{Maldacena} proposed as a conjecture that $IIB$ string theories in a background
$AdS_5\times S_5$ with  radius $l\sim l_s (g_s N)^{1/4}$ and N units of RR flux 
is equivalent to a four dimensional ordinary gauge theory in flat four-dimensional Minkowski
space, namely ${\cal N}=4$ super Yang-Mills with gauge group $SU(N)$ and coupling constant
$g=g_s^{1/2}$.
 \par
Although there is much supersymmetry in the problem and the kinematics largely determine correlators,
(in particular, the symmetry group $SO(2,4)\times SO(6)$ is realized as an isometry group on the
gravity side and as an $R$-symmetry group as well as conformal invariance on the gauge theory side)
this is not fully so \footnote{The only correlators that are completely determined
through symmetry are the two and three-point functions.}
and the conjecture has passed many tests in the semiclassical approximation
to string theory.
\par
 This is the first time that a precise holographic description of spacetime in terms of a (boundary) 
gauge theory is proposed and, as such it is of enormous potential interest. It has been conjectured 
by 't Hooft \cite{'tHooft} and further developed by Susskind \cite{Susskind} that there should be
much fewer degrees of freedom in quantum gravity than previously thought. The conjecture claims that
it should be enough with one degree of freedom per unit Planck surface in the two-timensional boundary
of the three-dimensional volume under study. The reason for that stems from an analysis of the 
Bekenstein-Hawking \cite{Bekenstein}\cite{Hawking} 
entropy associated to a black hole, given in terms of the
two-dimensional area $A$ \footnote{The area of the horizon for a Schwarzschild black hole is given by:
\be
A=\frac{8\pi G^2}{c^4}M^2
\ee
}
of the horizon by
\be
S=\frac{ c^3}{4 G\hbar}A.
\ee

This is  a deep result indeed, still not fully understood.
\par
It is true on the other hand that the Maldacena conjecture has only been checked for the time being 
in some corners of 
parameter space, namely when strings can be approximated by supergravity in the appropiate background.
The way it works \cite{Witten} is that the supergravity action corresponding to fields with prescribed
boundary values is related to gauge theory correlators of certain gauge invariant operators 
corresponding to the particular field studied:
\be
e^{- S_{sugra}[\Phi_i]}\bigg|_{\Phi_i|_{\partial AdS}=\phi_i}= 
<e^{\int {\cal O}_i\phi_i}>_{CFT}
\ee

 \section{Observational prospects}

In the long term, advances in the field, as in any other branch of physics will be determined 
by experiment. The prospects here are quite dim. It has been advertised \cite{Alfaro} that
as a consequence of loop quantum gravity 
\footnote{In string theory, with a  string scale of the order of the Planck mass, there is effectively
a minimal length, namely the self dual radius but then
 the corrections are of the type \cite{Myers}
\be
E^2 = (\vec{p})^2 + m^2 + m^2\sum_{n=1} c_n (\frac{m}{m_P})^n
\ee
much more difficult to observe experimentally}
anomalous dispersion relations of the form
\be
E^2 = (\vec{p})^2 + m^2 + E^2\sum_{n=1} c_n (\frac{E}{m_P})^n
\ee
could explain some strange facts on the cosmic ray spectrum. Although this is an interesting 
suggestion (cf.\cite{Coleman}) it is not specific to loop quantum gravity; noncommutative models 
make similar predictions as indeed does any theory with a fundamental scale. In spite of some optimism, it is not easy to perform specific 
experiments which
could discriminate between different quantum gravity alternatives. This should not by any means
be taken as an indication that the experiments themselves are not interesting. Nothing could be most
exciting that to discover deviations from the suposedly exact symmetries of Nature, and it is 
amazing that present observations already seemingly exclude some alternatives \cite{Myers}.
\par
On the string side, perhaps some effects related to specific stringy states, such as
the winding states could be experimentally verified ( cf. for example some suggestions in 
\cite{Alvarezz}). It has also been proposed that the string scale
could be lowered, from the Planck scale down to the TeV regime \cite{Antoniadis}.
It is difficult to really pinpoint what is exactly stringy about those models, and in particular,
all string predictions are difficult to disentangle from supersymmetric 
model predictions and rely heavily
on the mechanisms of supersymmetry breaking.

\section{Summary: the state of the art in quantum gravity}
In the loop approach one is working
 with  nice candidates for a quantum
  theory. The theories are interesting, probably related to topological field theories (\cite{Blau})
and background independence as well as diffeomorphism invariance are clearly implemented.
On the other hand, it is not clear that their low energy limit  is related to 
Einstein gravity.
    
\par 
 Strings start from a perturbative approach more familiar to a particle physicist.
However, they carry all the burden of supersymmetry and Kaluza-Klein. It has proved to be very
difficult to study nontrivial non-supersymmetric dynamics.
\par
Finally, and this applies to  all approaches, the holographic ideas seem intriguing; there are
many indications of a deep relationship between gravity and gauge theories.
\par
We would like to conclude by insisting on the fact that although there is not much we know for 
sure on quantum effects on the gravitational field, even
the few things we know are a big feat, given the difficulty to do physics 
without experiments.

\section*{Acknowledgments}
I wish like to thank Norma Mankoc-Borstnik and Holger Bech Nielsen for the oportunity to
participate in a wonderful meeting in the pleasant atmosphere of Portoroz.
I have benefited from discussions with J. Alfaro, J. Barrett, C. G\'omez, W. Kummer  and  E. Verdaguer.
This work ~~has been partially supported by the
European Commission (HPRN-CT-200-00148) and CICYT (Spain).

\appendix

\end{document}